\documentclass[a4paper]{article}
\usepackage[utf8]{inputenc}
\usepackage[english]{babel}
\usepackage[12pt]{extsizes}
\usepackage[centertags]{amsmath}
\usepackage{graphicx}
\usepackage{multirow}
\usepackage{hhline}
\usepackage{amssymb}
\usepackage{setspace}
\usepackage[left=3cm, right=20mm, top=2cm, bottom=2cm]{geometry}
\usepackage{indentfirst}

\usepackage{latex_nii_R}
\usepackage{here}

\begin{document}





%


\title{\large{\bf{Three-component St\"ackel model of the Galaxy based\\ on the
rotation curve from maser data}}}


\author{\normalsize{A.~O.Gromov\thanks{e-mail: granat08@yandex.ru}, \quad I.I. Nikiforov}}
\date{\normalsize{St. Petersburg State University, St. Petersburg, 198504 Russia}}







\maketitle


\begin{abstract}
A three-component St\"ackel model of the Galaxy, including the bulge, disk, and halo, is constructed.
Parameter estimates of the potential are obtained as a result of fitting the model rotation curve to
azimuthal velocities found from data on trigonometric parallaxes and spatial velocities of masers. The fitting
method takes into account the measurement and natural dispersions of azimuthal velocities and uses an algorithm
for excluding objects with excessive residuals. In order to obtain more uniform samples, the objects
were divided into two groups: masers associated with high-mass star forming regions and masers of other
types. A significant kinematic inhomogeneity of these groups was identified and taken into account: the azimuthal
velocity dispersion is $\sigma_{0,1}=4.3\pm 0.4$~km\,s$^{-1}$, in the first group and
$\sigma_{0,2}=15.2\pm1.3$~km\,s$^{-1}$ in the second.
After constructing the model of the Galactic-plane potential, it was generalized to the entire space under
the assumption of the existence of a third quadratic integral of motion. When reconstructing the Galactic
rotation curve in detail, the used algorithm gives an analytical expression for the St\"ackel potential, which significantly
simplifies the task of constructing the Galaxy's phase density model in the St\"ackel approximation.
In order to make the St\"ackel model more realistic, one needs to develop methods of direct account of data on
the vertical distribution of density in the Galaxy.

\end{abstract}




\section{\large{INTRODUCTION}}



St\"ackel models are widely discussed in papers
related in one way or another to constructing phase
models of star systems, including those of our Galaxy.
This is due to the possibility of the existence in such
models of a third integral of motion, quadratic in
velocity (in addition to the two classical integrals of
energy and areas)
\begin{equation}\label{I3}
I_3=(Rv_z-zv_R)^2+z^2v_{\lambda}^2+z_0^2\left(v_z^2-2\Phi^*\right)\,,
\end{equation}
where
$$
\displaystyle z_0^2\frac{\partial\Phi^*}{\partial R}=
z^2\frac{\partial\Phi}{\partial R}-Rz\frac{\partial\Phi}{\partial z}\,,\qquad
$$
$$
\displaystyle z_0^2\frac{\partial\Phi^*}{\partial z}=\left(R^2+z_0^2\right)
\frac{\partial\Phi}{\partial z}-Rz\frac{\partial\Phi}{\partial R}\,.
$$
Here $R$, $\lambda$, $z$~are the cylindrical coordinates, $\Phi$~is the
potential of the model considered, and $z_0$~is some constant
of the dimension of length.

Such an integral is widely applied in stellar dynamics,
for example, in the works of Kuzmin (1952), Kuzmin
and Malasidze (1987), Osipkov (1975) and many
other authors.

The velocity ellipsoid for most subsystems in the
Galaxy, determined from the proper motions and
radial velocities of the stars, is triaxial (see, e.g., the
data on the principal velocity dispersions of main
sequence stars with various $B-V$ color indices in a
monograph by Binney and Merrifield (1998)). One of
the ellipsoid axes coincides with the $v_{\lambda}$ axis. If the
phase density depends on three integrals of motion,
then this fact can be explained within the framework
of the stationary Galaxy theory. Additionally, for
models with three integrals of motion, the problem of
computing orbits can be solved in quadratures. Since
the radial velocity dispersion exceeds significantly the
vertical one not only for nearby stars but also for
objects representing a large interval of galactocentric
distances (e.g., Bond et al., 2010; Rastorguev et al.,
2017), hereinafter we shall assume for the entire Galaxy
the existence of a global third integral as a phase
density argument. Note that the choice of parameter $z_0$ (see Section 4), the possible variations of which represent
the difference in the local integrals for individual
stars, can be considered as a selection of some average $z_0$
for the latter, which even in this case allows
us to talk about some common quasi integral of
motion.

For practical purposes, one should consider only
the single-valued integrals of motion: if non-isolating
integrals become arguments of phase density, then it
becomes infinitely multivalued, which lacks physical
sense (see, e.g., the monograph by Ogorodnikov,
1958). H\'enon and Heiles (1964) have shown numerically
that when energy reaches some critical value,
ergodic areas are formed where the third integral
becomes non-isolating. Rodionov (1974) determined
that in order to avoid such \flqq  ergodic layers\frqq\  a sixfold
continuous differentiability of the potential at the center
of the system is needed.

The possibility of the existence of such third integrals
has also been considered for other galaxies (e.g.,
Binney et al., 1990; Merrifield, 1991).

The condition of the existence of the third quadratic
integral of motion (1) is written in the form of
\begin{equation}\label{6}
\displaystyle 3\left(z\frac{\partial\Phi}{\partial R}-R\frac{\partial\Phi}{\partial z}\right)-\left(R^2+z_0^2-z^2\right)\frac{\partial^2\Phi}{\partial R\partial z}
 +Rz\left(\frac{\partial^2\Phi}{\partial R^2}-\frac{\partial^2\Phi}{\partial z^2}\right)=0\,.
\end{equation}

For further computations it is convenient to switch
to elliptical coordinates: $\xi_1\in[1;\infty)$, $\xi_2\in[-1;1]$,
\begin{equation}
R=z_0\sqrt{\left(\xi^2_1-1\right)\left(1-\xi^2_2\right)}\,, \quad z=z_0\xi_1\xi_2.
\end{equation}
The limitation~\eqref{6} can then be written as
\begin{equation}\label{f5}
\frac{\partial^2}{\partial \xi_1 \partial \xi_2}\left[\left(\xi_1^2-\xi_2^2\right)\Phi\right]=0\,,
\end{equation}
and the potential, in turn, as follows:
\begin{equation}\label{f4}
\Phi=\frac{\varphi(\xi_1)-\varphi(\xi_2)}{\xi^2_1-\xi^2_2}\,,
\end{equation}
where $\varphi(\xi)$ is an arbitrary function.

Expression~\eqref{f5} is the condition of separating the
variables, which means that such potentials are separable,
i.e., they allow one to separate the variables in
the Hamilton–Jacobi equation. St\"ackel (1890) was the
first to introduce in his works a class of potentials that
satisfy~\eqref{f4}. They later became known as St\"ackel potentials
and were introduced into stellar dynamics by
Eddington (1915).

Another advantage of St\"ackel potentials in phase
modeling is the ability to describe models in angle--action
variables. Multiple expressions for
phase density with such variables are encountered in
the literature (e.g., Dehnen, 1999; Posti et al., 2015).
The algorithm for determining the actions and angles
in separable potentials (St\"ackel potentials are one type
of such potentials) was developed by J. Binney and his
team. This approach replaced the method of constructing
tori (Kaasalainen and Binney, 1994), which turned out to be inconvenient since it gives a dependence
of the phase variables on actions and angles and
not vice versa, as is required, and the method of adiabatic
invariants by Binney (2010), which was satisfactory
only for stars close to the equatorial plane.

To introduce the St\"ackel potential into models,
Sanders (2012), Binney (2012), Sanders and Binney
(2016) use the \flqq St\"ackel fudge\frqq\  algorithm, where the
$\varphi(\xi)$ function is an interpolation based on some number
of orbital points in the assumption that the non-St\"ackel potential in those points has the properties of
the St\"ackel one. However, this approach is approximate.
Additionally, the $z_0$ parameter changes from
orbit to orbit in such an approach. Due to this, one can
only talk of some local third integral, since the constancy
requirement for $z_0$
is key to keeping integral $I_3$. Which integral is the phase density argument in the
case of a non-constant $z_0$ is not quite clear.

Undoubtedly, studies within the framework of the
St\"ackel fudge have provided a large contribution to the
development of star system phase modeling methods.
However, since global integrals should, strictly speaking,
be phase density arguments, we suggest that an
alternative approach based on the $I_3$ integral with constant $z_0$, as another variant of fitting a more complex
reality, is also of interest. In contrast to the numerical
multi-step St\"ackel fudge algorithm, such an
approach is analytical and more simple. This paper is
dedicated to the attempt to implement this analytical
approach. The future will show which method will
prove to be more accurate and convenient.

The first step to construct a phase model is obtaining
the model of the potential that is consistent with
observations. Unfortunately, there are only a few studies
where St\"ackel potentials are constructed based on
observations. One of the first was the work of Satoh and
Miyamoto (1976), where a one-component model of
the Galaxy was determined by the data on Galactic
rotation and by the density in the solar vicinity.
However, that study uses only 18 objects with distances
up to 10 kpc. Additionally, the paper by Satoh and Miyamoto
(1976) adopts the value $\rho_{\odot}=0.148$~$M_{\odot}\,pc^{-3}$,
which does not correspond to the contemporary results (Bland-Hawthorn and Gerhard, 2016; Loktin and
Marsakov, 2009).

In a later work, Famaey and Dejonghe (2003) constructed
St\"ackel models of the Galaxy based on the
derivatives, mainly local, of the dynamical characteristics
(local circular velocity, flat rotation curve, Oort
constants and others). Such models are only representative;
in future, as Famaey and Dejonghe (2003)
themselves note, more extensive kinematic data
should be used, but such work has not been undertaken.

Formally, the models presented in Famaey and
Dejonghe (2003) are three-component, however, the
third component only appears in them due to an addition,
mainly for demonstrative purposes, of a thick disk component to the spheroid and (thin) disk, the
dynamic contribution of which is small compared to
the thin disk (Bland-Hawthorn and Gerhard, 2016).

We should also note the paper by Binney and Wong
(2017), where the \flqq St\"ackel fudge\frqq\ algorithm is applied
to data on Galactic globular clusters and a phase
model is constructed for that cluster system. However,
as was mentioned above, using the \flqq St\"ackel fudge\frqq\
algorithm gives only an approximation of the St\"ackel
model.

This work, where we implement an analytical
approach to constructing the St\"ackel model for the
Galaxy, is the first in a series of papers which as the
end result aims to construct a phase model of the Galaxy
which agrees with the large amount of information
obtained from observations. Based on kinematic data
on masers with trigonometric parallaxes we find the
optimal parameters for the model potential in the
Galactic plane, which we then generalize to three
dimensions in the assumption of the existence of a
third quadratic integral of motion. As a result we construct
a three-component (halo, disk, bulge) St\"ackel
model of the Galaxy based on the current rotation
data; we then discuss the feasibility of this model.

\section{\large{METHOD}}

St\"ackel models of the Galaxy constructed in the
papers mentioned above are merely approximate (representative
or roughly St\"ackel). The problem may be
solved by a method proposed by Rodionov (1974),
which allows one to generalize the equatorial plane
potential to the entire space in the St\"ackel way, obtaining
an analytical expression as a result. Since no limitations
are applied in this case to the equatorial plane
potential, any model may be used to construct the
St\"ackel model, for example, one obtained by approximating
the data on the rotation of the \flqq cool\frqq\ Galactic
subsystem encompassing a large interval of distances $R$. In particular, one can use maser data, and, as a possibility,
the Gaia catalog.


If the potential in set in the equatorial plane, $\varphi(\xi)$ is determined as (Rodionov, 1974):
\begin{equation}\label{f8}
\varphi(\xi)=\xi^2\;\Phi\left(R=z_0\sqrt{|\xi^2-1|},z=0\right)\,.
\end{equation}
Such a choice of function $\varphi(\xi)$ facilitates the fulfillment
of the following conditions:
\begin{enumerate}

\item[1)]$\varphi(0)=0$;

\item[2)]$\varphi(1)=\Phi_0$, where $\Phi_0$~--- is the potential in the center of the model;

\item[3)]when $\xi\to \infty$ $\displaystyle \frac{\varphi(\xi)}{\xi} \to \frac{GM}{z_0}$, where
$G$~--- is the gravitational constant and, $M$~--- is the total mass.


\end{enumerate}

The author of the method (Rodionov, 1974) also
proposed ways to estimate the function $\varphi(\xi)$, if the
potential is defined on the $z$ and in an arbitrary column $(R_*,z)$, where $R_*=\text{const}$.

We used this method earlier to construct one- and
two-component St\"ackel test models of the Galaxy
(Gromov et al., 2015, 2016). We used the data on neutral
hydrogen rotation as well as those on masers. A
good agreement between model rotation curves and
observed data was obtained, and also between model
values and estimates of density in the solar vicinity,
mass in a sphere of 50 kpc radius and other dynamic
characteristics.

In this work we use the same method to construct a more realistic three-component St\"ackel model of
the Galaxy with the classical composition: halo, disk, and bulge. Since the goal is to determine the applicability
of Rodionov’s method in a separate representation of the main dynamic components of the Galaxy in
the model, further refinement of the model (for instance, adding a thick disk) seems excessive at this
stage.

We chose a quasi-isothermal potential proposed in
Kuzmin et al. (1986) to describe the halo:
\begin{equation}
\Phi_1(R,0)=\Phi_{0,1}\ln\left(1+\frac\beta{w(R)}\right),
\end{equation}
where the~$w(R)$ function is determined as:
\begin{equation*}
w^2(R)=1+\kappa_1^2\,R^2\,.
\end{equation*}
The disk is presented as a general isochronous
potential (Kuzmin and Malasidze, 1969):
\begin{equation}
\Phi_2(R,0)=\Phi_{0,2}\,\frac{\alpha}{\left(\alpha-1\right)+\sqrt{1+\kappa_2^2 R^2}}\,.
\end{equation}
To describe the central bulge we used the Hernquist potential (Hernquist, 1990):
\begin{equation}
\Phi_3(R,0)=\Phi_{0,3}\,\frac{1}{R+\kappa_3}\,.
\end{equation}

According to \eqref{f8} the function  $\varphi_i(\xi)$ for individual components has the following form:
\begin{gather}\label{f13}
\varphi_1(\xi)=\xi^2\Phi_{0,1}\ln\left(1+\frac\beta
{\sqrt{1+\kappa_1^2z_0^2|\xi^2-1|}}\right),\\
\varphi_2(\xi)=\xi^2\Phi_{0,2}\,\frac{\alpha}{\left(\alpha-1\right)+\sqrt{1+\kappa_2^2z_0^2|\xi^2-1| }},\\
\varphi_3(\xi)=\xi^2\Phi_{0,3}\,\frac{1}{\sqrt{z_0^2|\xi^2-1|}+\kappa_3}.
\end{gather}
As a result, the final expression for the St\"ackel potential
function
\begin{equation}\label{f16}
\varphi(\xi)=\varphi_1(\xi)+\varphi_2(\xi)+\varphi_3(\xi)
\end{equation}
is analytical, which allows us to find the function value
and, as a consequence, the values of the action variables
in the entire region of defined phase coordinates, and not in isolated points.
The values of $\varphi(\xi)$ are determined
exactly here, and the accuracy of determining
the action variables that are expressed through complex
integrals is limited only by the accuracy of the
numerical methods used to find the latter.

\section{\large{OBSERVED DATA}}

In this work we use data on masers located in star forming regions. VLBI observations allow one to
obtain accurate estimates of trigonometric parallaxes and proper motions, including even for maser sources
that are rather distant from the Sun ($r< 10$~kpc) (Nikiforov and Veselova, 2018).

We used the catalog from Rastorguev et al. (2017) as the main maser database, which includes 103 high-mass
star forming regions (HMSFRs) from the homogeneous catalog of Reid et al. (2014) and 38 additional
maser sources of which, as determined by Nikiforov and Veselova (2018), nine belong to the HMSFR class
(see Table 1), and the remaining ones are of other types (hereafter non-HMSFR masers). Furthermore, we
added another 3 masers with full data: Sh2-76EMM1, Sh2-76EMM2
(G040.44+02.45) and AFGL 5142 (G174.20-00.07); they all belong to the non-HMSFR group. The
total sample thus consists of 144 objects

\begin{table}[]
\begin{center}
\caption{List of additional maser sources from the catalog of
Rastorguev et al. (2017), which belong to high-mass starforming
regions (HMSFRs) according to Nikiforov and
Veselova (2018)}
\label{Tab5}
\medskip
\begin{tabular}{l|l}
\hline
G170.66${-}$00.25 & G059.83+00.67 \\
G213.70${-}$12.60 & G071.52${-}$00.38 \\
G054.10${-}$00.08 & G305.200+0.019 \\
G058.77+00.64 & G305.202+0.208 \\
G059.47${-}$00.18 & \\
\hline
\end{tabular}
\end{center}
\end{table}

As was shown by Nikiforov and Veselova (2018), the measurement uncertainty for non-HMSFR maser
parallaxes is on average higher than for the HMSFR masers. Additionally, unlike the HMSFR sources
belonging to the same class of objects, non-HMSFR masers belong to 14 different classes (Nikiforov and
Veselova, 2018). This fact allows us to assume that the sample uniting these objects is kinematically inhomogeneous.
We therefore differentiated between HMSFR (112 objects) and non-HMSFR (32 objects)
masers when processing data in the general case. In particular, dynamic modeling was carried out separately
for the more homogeneous HMSFR group and for the joint (HMSFR + non-HMSFR) sample. We
did not consider the non-HMSFR sample separately, since its volume is insufficient for completing the outlined
task. As we show below, HMSFR and non-HMSFR masers are indeed kinematically inhomogeneous
with respect to each other.


\section{\large{PARAMETER OPTIMIZATION}}

The parameters of the model potential described in
Section 2 were estimated using the nonlinear least
squares method. The residual for an individual object
was the difference between the model circular velocity $\theta_C(R_i)$, where $R_i$ is the galactoaxial distance of the said
object, and azimuthal velocity $\theta_i$, computed by the
measured position and three velocity components of the object (see details in Gromov et al. (2016)\footnote{In item 7 of the Appendix in Gromov et al. (2016) the correct formula is
$\sin\beta=\frac{r\cos b}{R}\sin l$.}).
Since the natural azimuthal velocity dispersion of masers
was considered as a-priori unknown, at the initial
stage we minimized the objective function
\begin{equation}\label{L2}
L^2=\sum_{i=1}^{N}p_i\left[\theta_i-\theta_C(R_i)\right]^2,
\end{equation}
where
\begin{equation}\label{pi_sigma2i}
\displaystyle p_i=\frac{1}{\sigma^2_i}
\end{equation}
is the weight of the $i$th object that accounts only for the
measurement uncertainty $\sigma_i$
of velocity $\theta_i$ (see Gromov et al. (2016)); $\displaystyle \theta_C^2=-R\,\frac{d\Phi}{dR}$,
$\Phi(R,0)=\Phi_1(R,0)+\Phi_2(R,0)+\Phi_3(R,0)$; $N$ is the volume of the sample considered.

\begin{figure*}
\includegraphics[scale=0.8]{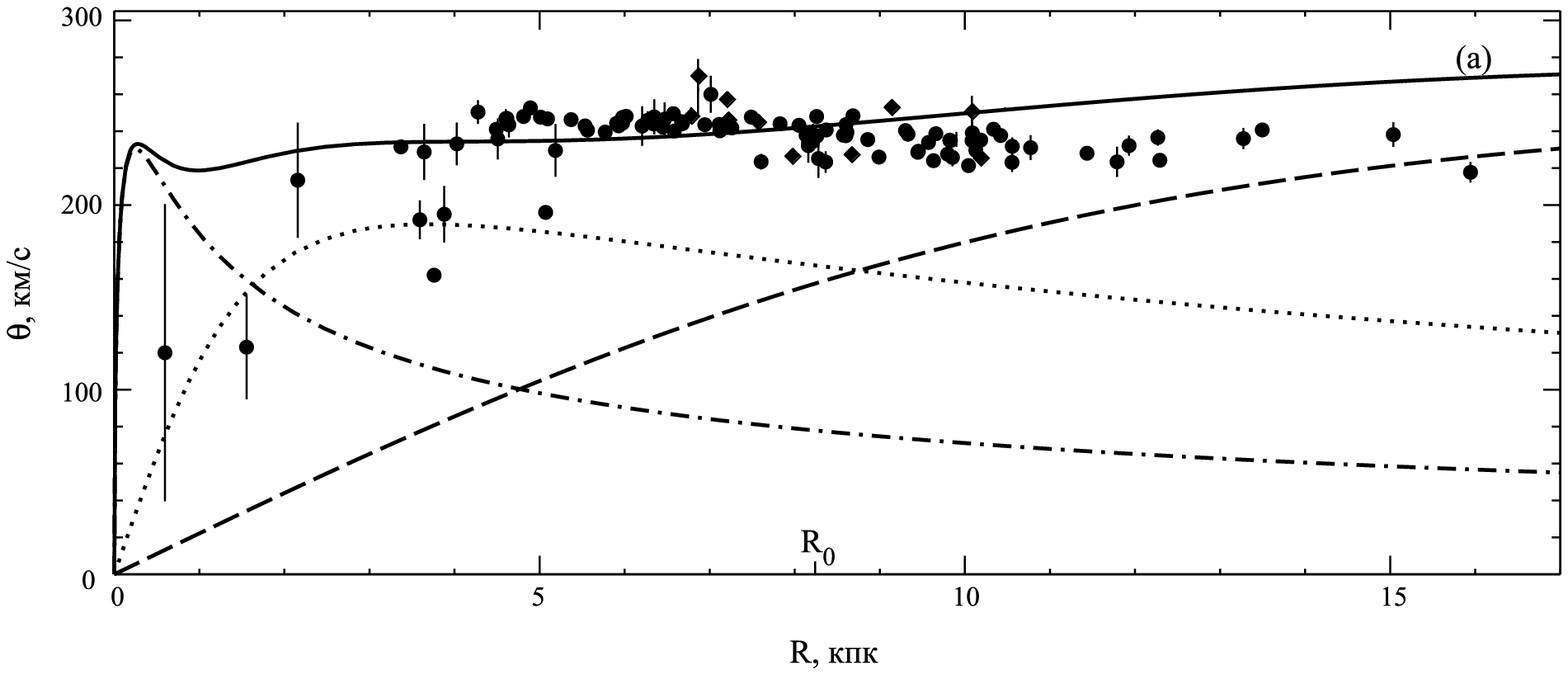}
\includegraphics[scale=0.8]{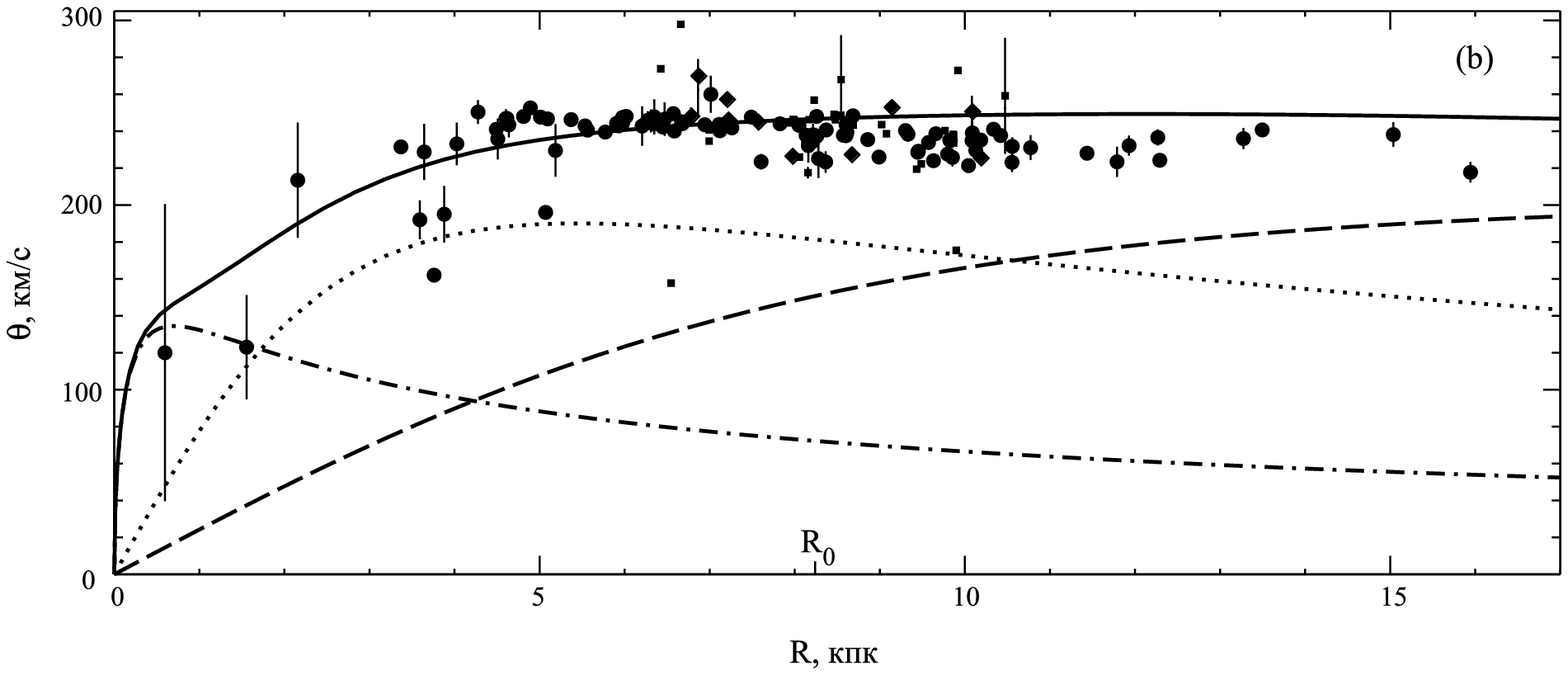}
\caption{Model circular velocity curves, obtained without
account of the natural dispersion of azimuthal velocities:
from HMSFR masers (a), from a joint HMSFR + non-
HMSFR sample (b). The solid curve is the three-component
model of the Galaxy; the dashed curve shows the
quasi-isothermal component representing the halo; the
dotted line is the general isochronous component that
describes the disk; the dashed-and-dotted curve is the
Hernquist component representing the bulge. Also shown
are the maser azimuthal velocities found from their measured
parallaxes, radial velocities, and proper motions.
The circles show the HMSFR masers from the catalog of
Reid et al. (2014); rhombi show the HMSFR masers from
the additional list in Rastorguev et al. (2017); small squares
show the non-HMSFR masers from the list of Rastorguev
et al. (2017) (they have little effect on the solution). The
bars show the initial data measurement errors. The
absence of a bar means that it is smaller than the size of the
symbol representing the object. The horizontal axis shows
the distance of the Sun from the center of the Galaxy
adopted in this work, $R_0=8.34$~kpc Reid et al. (2014).}
\vspace{30pt}
\label{Fig1}
\end{figure*}

The parameter estimates for HMSFR masers and for
those of the joint sample, obtained by minimizing ~\eqref{L2} are presented in Table~\ref{Tab1}, the corresponding model
rotation curves are fitted to the observed data in Fig~\ref{Fig1}. Since some computation versions in this work have
limiting parameter $\beta$ values~---
$\beta\rightarrow\infty$ (in this case the
quasi-isothermal potential becomes the Schuster--Plummer potential),--- the associated parameter $q=\displaystyle\frac{\beta}{\beta+1}\in[0;1)\,$ was estimated instead.

\begin{table}[]
\begin{center}
\caption{Solution without account for the natural azimuthal
velocity dispersion}
\label{Tab1}
\medskip
\begin{tabular}{c|c|c}
\hline
Characteristic & HMSFRs & HMSFRs + \\
& & non-HMSFRs \\
\hline
$N$ & 112 & 144\\
$q$ & $0.6748\pm 0.0073$ & $0.8421\pm 0.0022$ \\
$\kappa_1$, kpc$^{-1}$ & $0.0611\pm 0.0013$ & $0.2888\pm 0.0035$ \\
$\Phi_{0,1}$,~km$^2$\,s$^{-2}$ & $305.5\pm 6.5$ & $262.8\pm 1.1$ \\
$\alpha$ & $1.214\pm 0.023$ & $1.240\pm 0.019$ \\
$\kappa_2$,~kpc$^{-1}$ & $0.533\pm 0.015$ & $0.10102\pm 0.00029$ \\
$\Phi_{0,2}$,~km$^2$\,s$^{-2}$ & $400.6\pm 5.0$ & $310.83\pm 0.97$ \\
$\Phi_{0,3}$,~km$^2$\,s$^{-2}$ & $230.4\pm 14.3$ & $233.7\pm 3.7$ \\
$\kappa_3$,~kpc & $0.25\pm 0.12$ & $0.70\pm 0.24$ \\
$L^2/N_\text{free}$& $236$ & $713$ \\
\hline
\end{tabular}
\end{center}
\end{table}

When minimizing~\eqref{L2} with weights~\eqref{pi_sigma2i} the values of $L^2/N_\text{free}$ (reduced statistics $\chi^2$),
where $N_\text{free}=N-M$ is the number of degrees of freedom, $M$ is the number of parameters to be determined,
turned out to be much larger than unity (Table~\ref{Tab1}). This
implies that the deviations of azimuthal velocities~$\theta_i$ of masers from the model cannot be explained only by
measurement errors, and therefore, the natural velocity dispersion has to be taken into account even for
such the \flqq cool\frqq\ disk subsystem as masers (in accordance with the results of Rastorguev et al., 2017).

Therefore, at the second stage, the natural dispersion $\sigma_0$ of maser azimuthal velocities was introduced
into the objective function as an unknown parameter
by writing the weight coefficients as:
\begin{equation}\label{pi_sigma02}
p_i=\left(\sigma_i^2+\sigma_0^2\right)^{-1}\,,
\end{equation}
where $\sigma_i$~--- is the measurement uncertainty $\theta_i$.

The value $\sigma_0$ was estimated in an iterative procedure, each step of which had fixed parameters of the
potential determined in the previous step, and the natural dispersion was determined from the following
equation:
\begin{equation}\label{sigma02}
L^2\left(\sigma_0^2\right)=N_\text{free}\,.
\end{equation}
Using the~$\sigma_0$ obtained at this stage, the parameters
of the potential were determined anew as a result of minimizing
function~\eqref{L2} with the system of weights~\eqref{pi_sigma02}. The found parameter estimates were used in the next
step to find a new approximation of $\sigma_0$. In all cases the
procedure converged in three steps. As an initial
approximation we took the result obtained with the
system of weights~\eqref{pi_sigma2i}.

To verify and account for the kinematic inhomogeneity
of the non-HMSFR maser group with respect to the HMSFR group, in addition to the natural dispersion,
common for the joint HMSFR + non-HMSFR sample (let us call this variant \flqq approach 1\frqq), we also
estimated this dispersion for the HMSFR group, $\sigma_{0,1}$, and the non-HMSFR group, $\sigma_{0,2}$, individually (let us call this variant \flqq approach 2\frqq). In the latter case we
used the weight coefficients $p_{i,1}=\left(\sigma_i^2+\sigma_{0,1}^2\right)^{-1}$ for the HMSFR objects, and $p_{i,2}=\left(\sigma_i^2+\sigma_{0,2}^2\right)^{-1}$  for non-HMSFR objects.

In the case of two natural dispersions equation~\eqref{sigma02} can be written as:
\begin{equation*}\notag
L^2\left(\sigma_{0,1}^2,\sigma_{0,2}^2\right)=\sum_{i=1}^{N_1}\frac{\left[\theta_i-\theta_C(R_i)\right]^2}{\sigma_i^2+\sigma_{0,1}^2}+
\sum_{i=N_1+1}^{N}\frac{\left[\theta_i-\theta_C(R_i)\right]^2}{\sigma_i^2+\sigma_{0,2}^2}=N_\text{free}\,,
\end{equation*}
where the first sum is taken by the HMSFR objects, and the second --- by the non-HMSFR objects. If the
dispersions in the denominators of the two sums are correct, the contribution of each of them into the total
result is asymptotically equal to the fraction of each group of objects in the joint sample. This gives the
equations for determining $\sigma_{0,1}$ and $\sigma_{0,2}$:
\begin{equation}
\sum_{i=1}^{N_1}p_{i,1}\left[\theta_i-\theta_C(R_i)\right]^2=\frac{N_1}{N}\,N_\text{free}\,,
\end{equation}
\begin{equation}
\sum_{i=N_1+1}^{N}p_{i,2}\left[\theta_i-\theta_C(R_i)\right]^2=\frac{N_2}{N}\,N_\text{free}\,,
\end{equation}
where $N_1$ and $N_2$ are the numbers of masers in HMSFR and non-HMSFR groups respectively, $N=N_1+N_2$;
for the total sample $N_1=112$, $N_2=32$, $N=144$. The $\sigma_{0,1}$ and $\sigma_{0,2}$
values were obtained as a result of an iterative procedure similar to the one
described above.

\begin{figure*}
\includegraphics[scale=0.8]{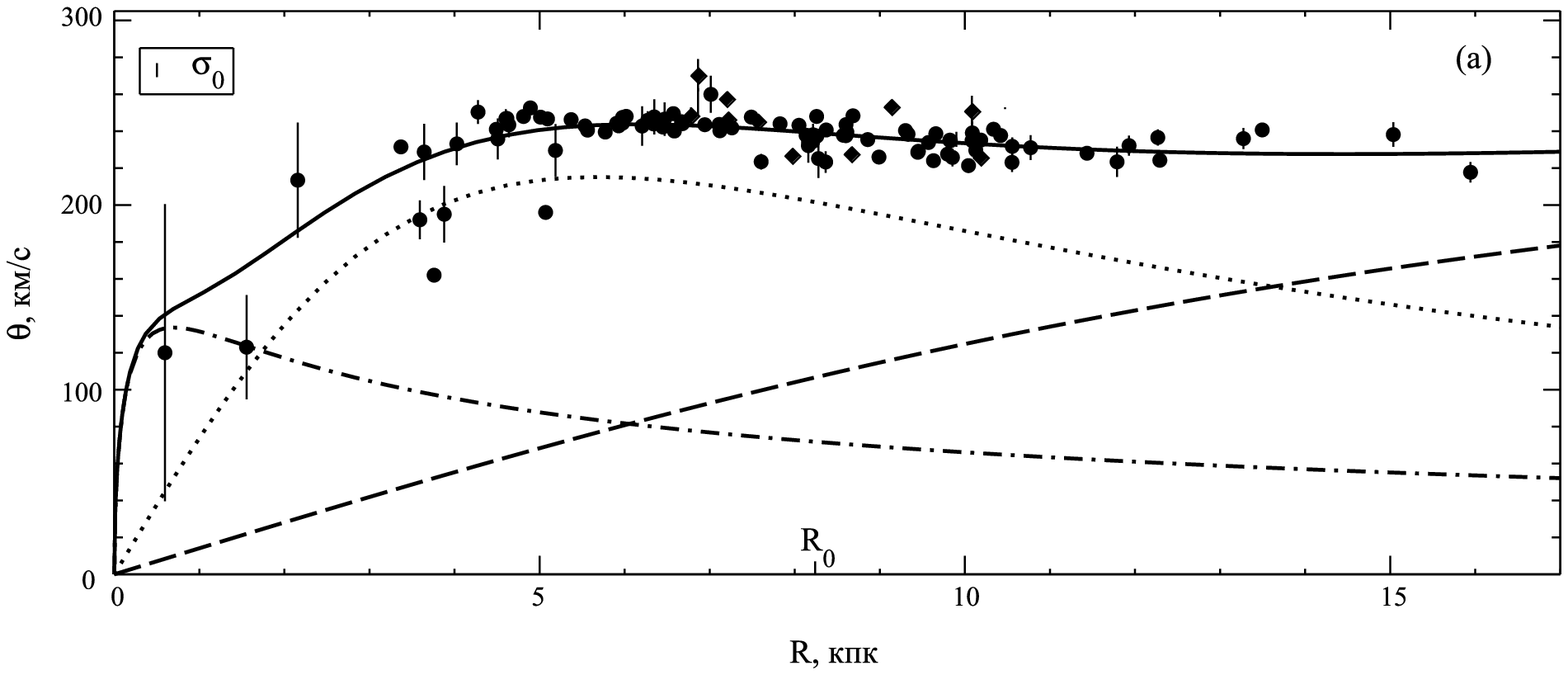}
\includegraphics[scale=0.8]{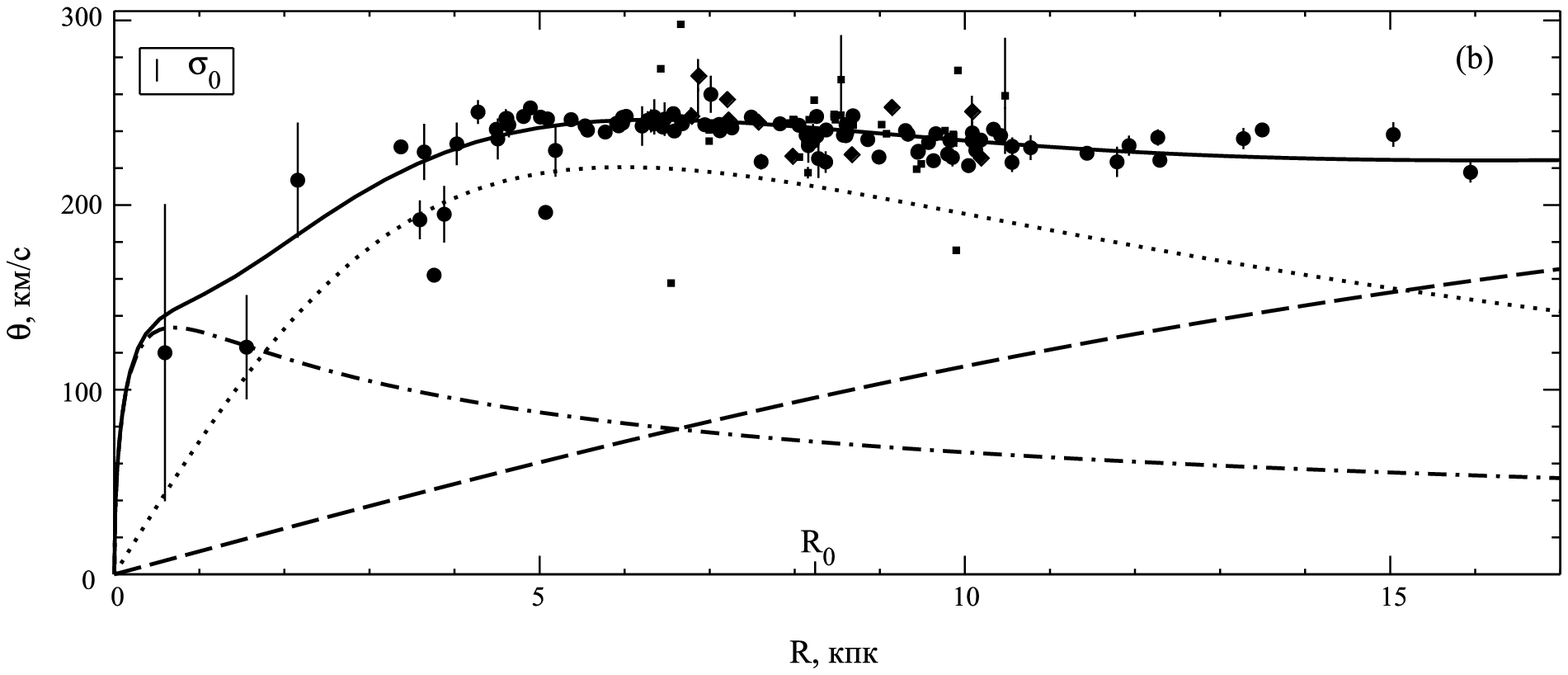}
\includegraphics[scale=0.8]{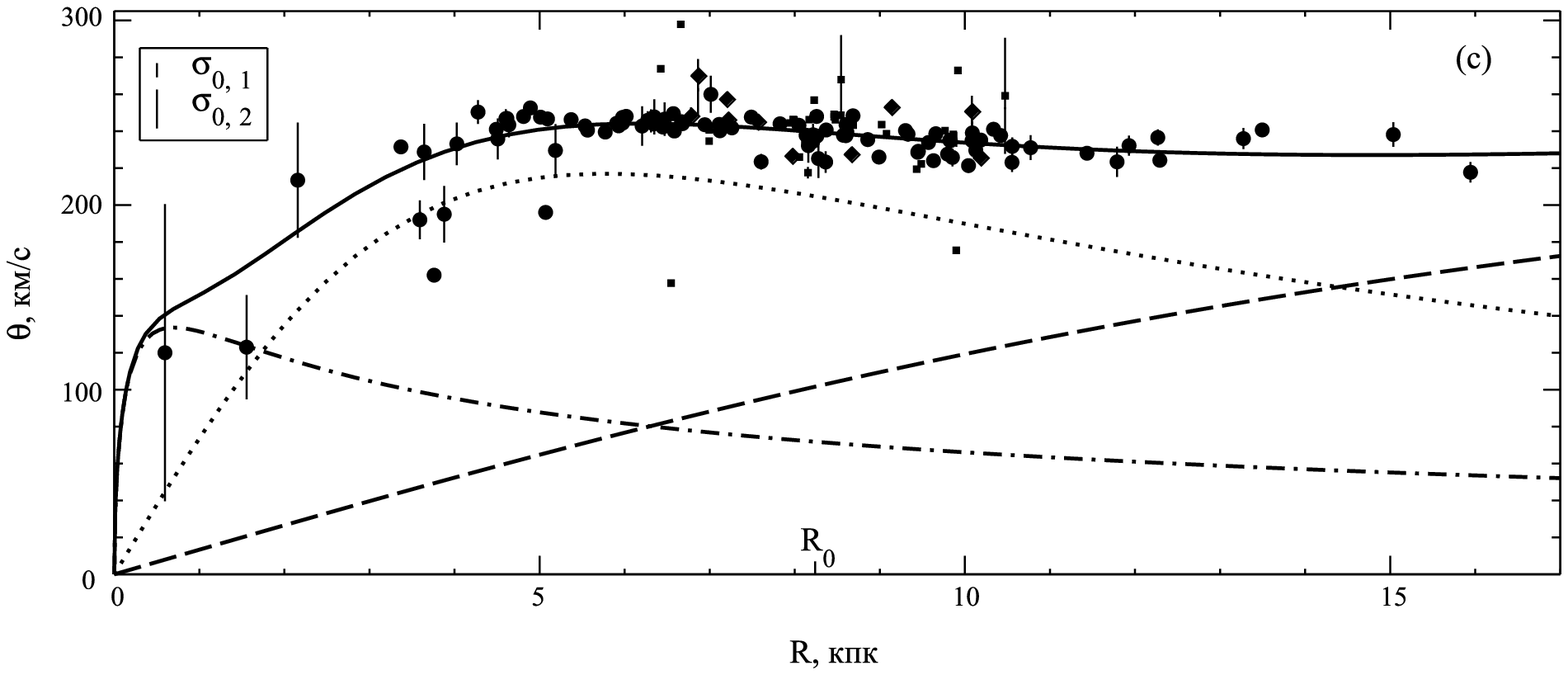}
\caption{Same as in Fig.~\ref{Fig1}, but with account for the natural
dispersion of azimuthal velocities: for the HMSFR sample
(a), for the joint HMSFR + non-HMSFR sample with the
same (b), and with different (c) natural dispersions for the
HMSFR and non-HMSFR types. The top left parts of the
panels show (in vertical axis scale) the bars of root-meansquare
deviations corresponding to the derived natural dispersions.}
\label{Fig2}
\end{figure*}

The results obtained at the second stage are presented in Table~\ref{Tab2} and Fig.~\ref{Fig2}.
\begin{table*}[]
\begin{center}
\caption{Solution with account for the natural azimuthal velocity dispersions}
\label{Tab2}
\medskip
\begin{tabular}{c|c|c|c}
\hline
Characteristic & HMSFRs & HMSFRs + non-HMSFRs & HMSFRs + non-HMSFRs \\
& & (approach 1) & (approach 2) \\
\hline
$N$ & 112 & 144 & 144\\
$\sigma_0$ or $\sigma_{0,1}$, km\,s$^{-1}$ & $7.67\pm0.74$ & $12.1\pm 1.0$ & $7.59\pm0.65$ \\
$\sigma_{0,2}$, km\,s$^{-1}$ & & & $20.9\pm1.8$ \\
$q$ & $1_{-0.035}$ & $1_{-0.053}$ & $1_{-0.035}$ \\
$\kappa_1$, kpc$^{-1}$ & $0.0773\pm 0.0026$ & $0.0734\pm 0.0031$ & $0.0956\pm 0.0032$ \\
$\Phi_{0,1}$, km$^2$\,s$^{-2}$ & $265.2\pm 9.4$ & $264.9\pm 15.1$ & $265.0\pm 10.2$ \\
$\alpha$ & $0.1437\pm 0.0041$ & $0.1148\pm 0.0048$ & $0.1651\pm 0.0062$ \\
$\kappa_2$, kpc$^{-1}$ & $0.0533\pm 0.0013$ & $0.04686\pm 0.0016$ & $0.0504\pm 0.0012$ \\
$\Phi_{0,2}$, km$^2$\,s$^{-2}$ & $311.2\pm 3.2$ & $318.4\pm 4.1$ & $317.4\pm 3.2$ \\
$\Phi_{0,3}$, km$^2$\,s$^{-2}$ & $224.5\pm 15.6$ & $223.4\pm 20.1$ & $223.8\pm 15.7$ \\
$\kappa_3$, kpc & $0.71\pm 0.32$ & $0.72\pm 0.34$ & $0.76\pm 0.31$ \\
\hline
\end{tabular}
\end{center}
\end{table*}

The solution without account for the natural azimuthal velocity dispersion was unstable with respect
to the sample composition (see the results for two samples in Table~\ref{Tab1} and in Fig.~\ref{Fig1}), obviously, due to the
exceedingly strong influence of objects with a high ratio of the absolute residual value to the (very small)
measurement error (examples are evident in Fig.~\ref{Fig1}). Introducing the natural dispersion has, for the most
part, eliminated this effect and stabilized the solution: all three results turned out to be close (Table~\ref{Tab2},
Fig.~\ref{Fig2}). Only the natural dispersions differ, and significantly:
adding non-HMSFR masers to the sample increases $\sigma_0$ by a factor of 1.5; direct dispersion measurements
for the two maser groups show that $\sigma_{0,2}$ is multiple
times larger than~$\sigma_{0,1}$. These facts imply a kinematic
inhomogeneity of the joint sample and justify the
introduction of two dispersions.

An analysis of the residuals of the derived solutions has shown that even after taking into account the natural
dispersion, the sample still contains objects with data outliers, which can substantially shift the solution.
To eliminate objects with excessive residuals at the third and final stage, we used the following algorithm
with a flexible boundary for criterion statistics (Nikiforov, 2012). We searched for objects for which
\begin{equation}\label{22}
\frac{|\theta_i-\theta_C(R_i)|}{\sqrt{\sigma_i^2+\sigma_{0}^2}}>k,
\end{equation}
where $k$ is determined from the equation
\begin{equation*}
\left[1-\psi(k)\right]N_\text{free}=1,
\end{equation*}
$\psi(z)=\displaystyle\sqrt{\frac{2}{\pi}}\int^z_0 e^{-\displaystyle\frac12\, t^2}dt$
is the probability integral and $\sigma_{0}$ is the corresponding
natural dispersion, i.e. $\sigma_0$, $\sigma_{0,1}$ or $\sigma_{0,2}$ depending on
the approach and object. Then out of $L$ objects that
satisfied the condition~\eqref{22}, $L-L'$ object were excluded, where $L'=3$  at the stage of determining  $\sigma_0$, $\sigma_{0,1}$, $\sigma_{0,2}$ or $L'=1$ at the final stage of estimating model
parameters with the derived (fixed) natural dispersions
(see Nikiforov, 2012). Objects with absolute
residual values exceeding $k_{0.05}$, i.e., the root of equation
\begin{equation*}
\left[1-\psi(k_{0.05})\right]N_\text{free}=0.05,
\end{equation*}
were also excluded, if found among the remaining objects.

After each iteration of exclusions, the problems of parameter optimization, natural dispersion determination,
and finding possible outliers in the data were solved anew until no object was excluded during the
next iteration.

The object G110.19+02.47 with a negative residual~$<3\sigma$, which is not formally excessive according the
criteria listed above, was also forcefully excluded within the framework of approach 2.
However, it is isolated from the array of other non-HMSFR masers in the histogram of the distribution of relative residuals.
This is the only object the class of which could not be determined from the original work of Chibueze
et al. (2014). These authors also note the low azimuthal velocity of this object, linking it with the local
kinematic anomaly, which does not characterize the Galaxy as a whole, of the Perseus arm in the region of
the rotation curve $R\sim 9$~kpc. Additionally, including this object in the sample noticeably
increases $\sigma_{0,2}$ from $15.2\pm1.3$ (final estimate) to $19.6\pm1.7$~km/s. Maser G110.19+02.47 was therefore
excluded as a possibly anomalous object.

\begin{figure*}
\includegraphics[scale=0.8]{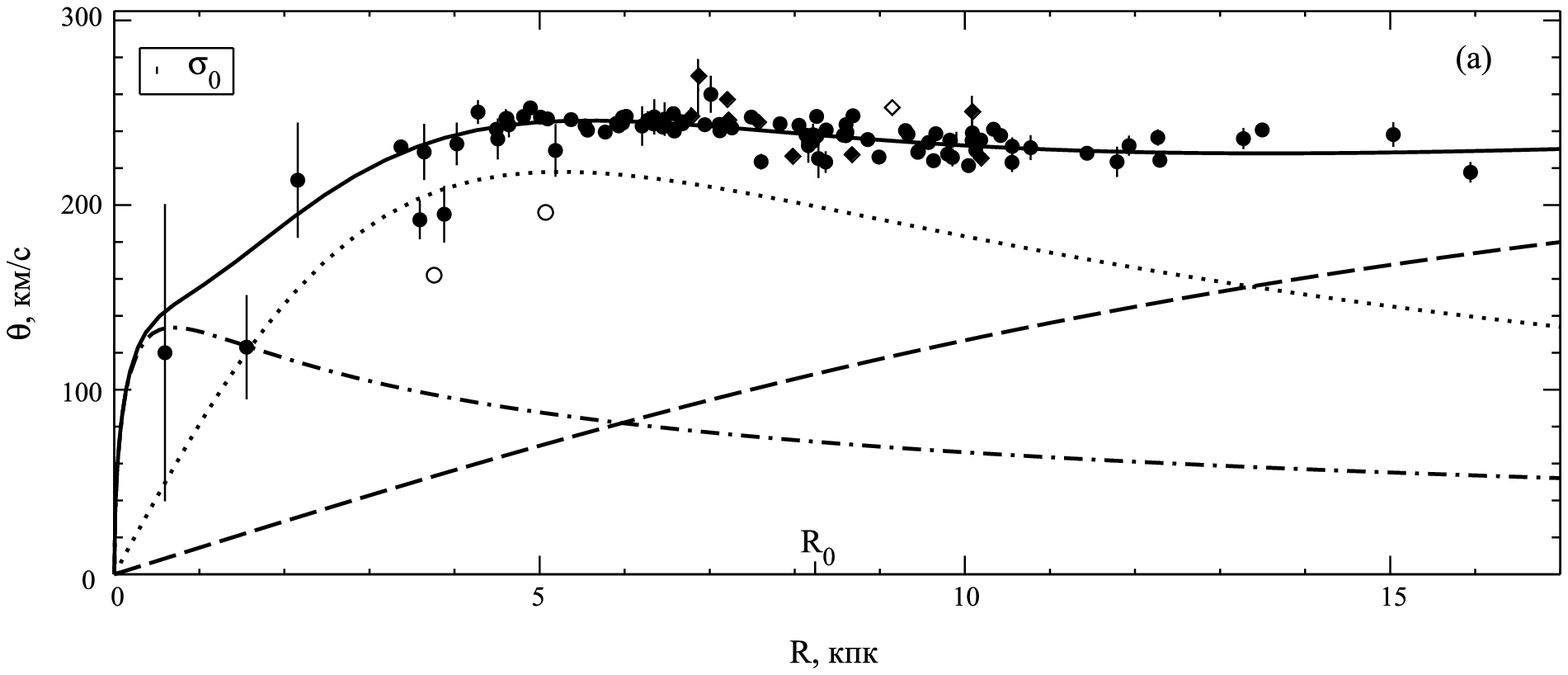}
\includegraphics[scale=0.8]{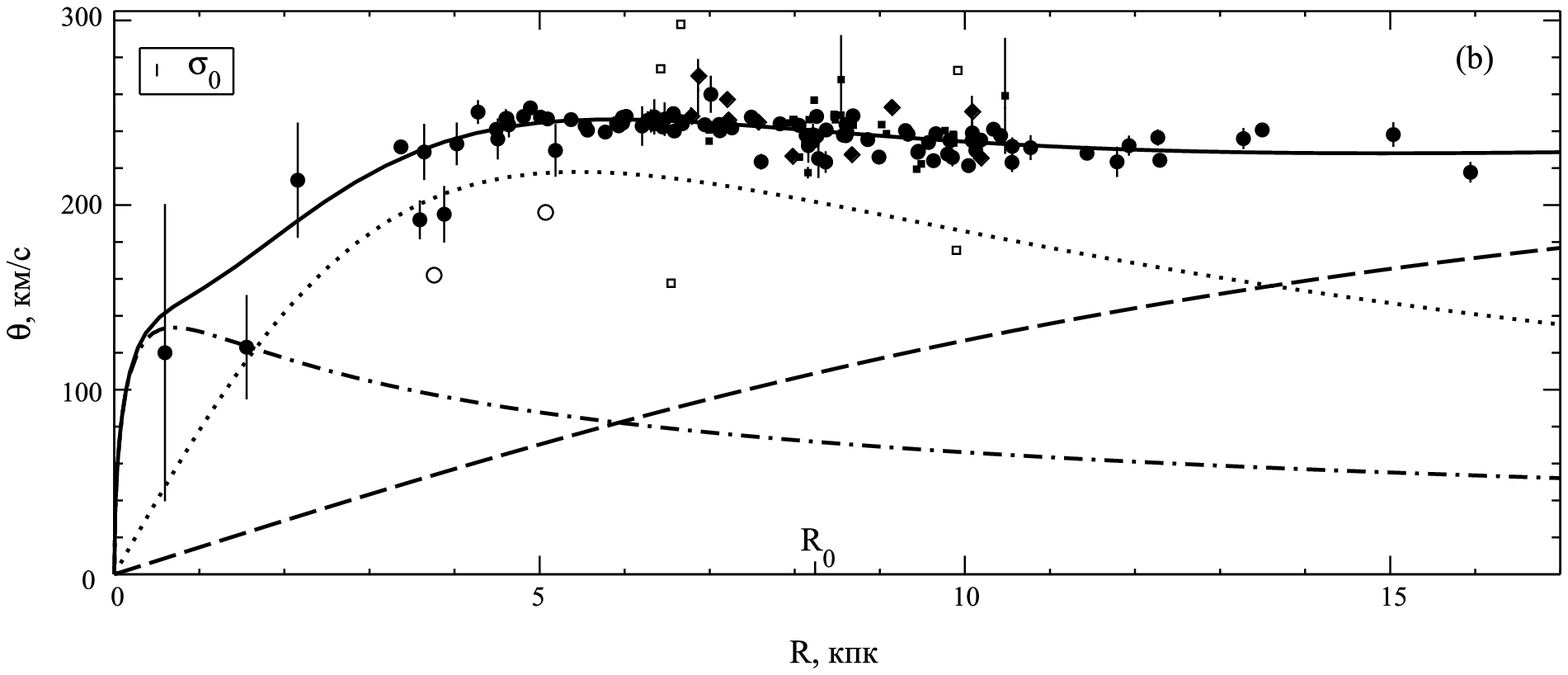}
\includegraphics[scale=0.8]{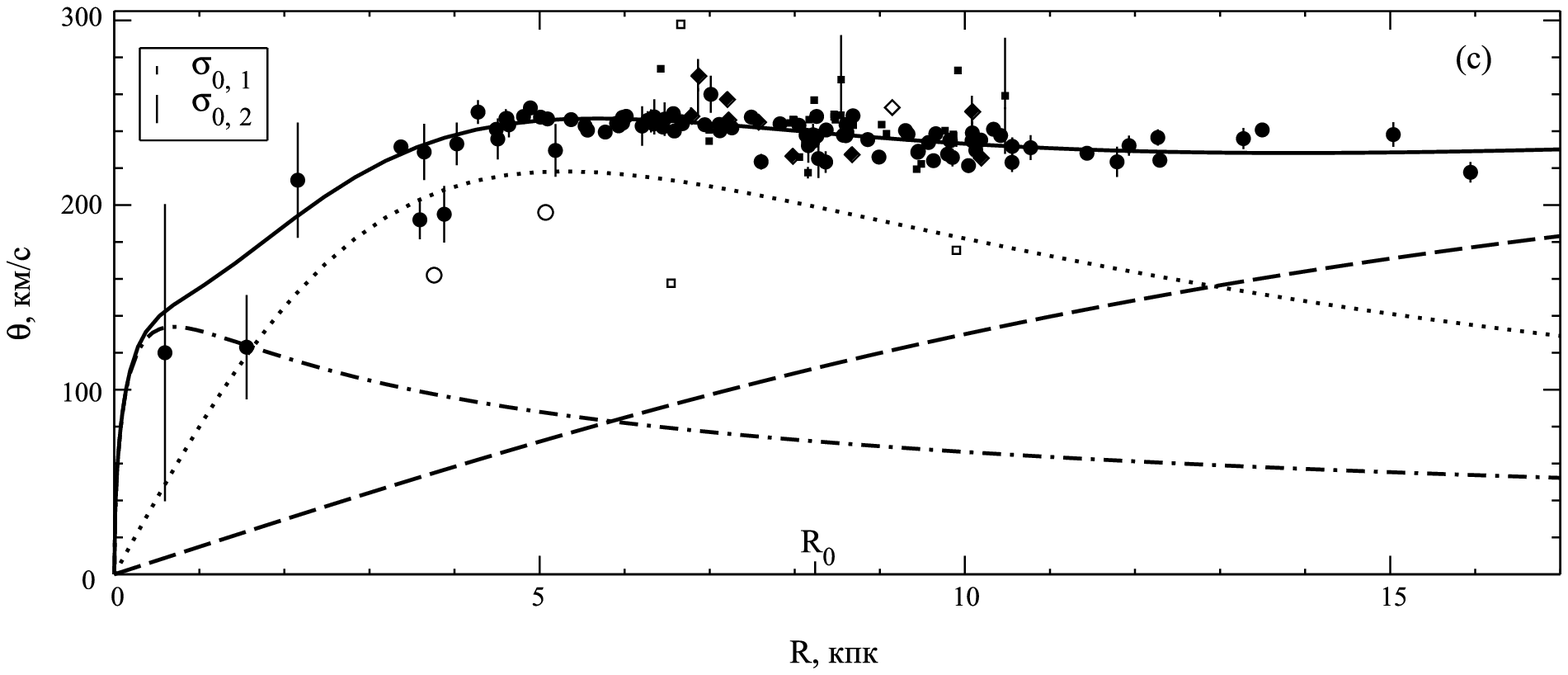}
\caption{Same as in Fig.~\ref{Fig2}, but after excluding from consideration
the masers with data outliers and the anomalous
object. The open symbols show the discarded objects.}
\label{Fig3}
\end{figure*}

The list of excluded objects for the samples and
approaches considered is shown in Table~\ref{Tab3}. The final
model parameter optimization results from the maser
data are presented in Table~\ref{Tab4} and in Fig.~\ref{Fig3}. Model
component masses within a 50~kpc radius sphere are as
follows: $M_{\text{bulge}}=1.1\times
10^{10}~M_{\odot}$, $M_\text{disk}=8.7\times 10^{10}~M_{\odot}$, $M_\text{halo}=7.2\times
10^{11}~M_{\odot}$.

\begin{table}[]
\begin{center}
\caption{Lists of objects excluded from consideration in
order of their exclusion during the final estimation of model
parameters ($L'=1$). The critical values of $k$ and
$k_{0.05}$ are
shown for the final sample volumes}
\label{Tab3}
\medskip
\begin{tabular}{l|c}
\hline
\multicolumn{1}{c|}{Object} &
$\rule{0em}{1.7em}\displaystyle\frac{|\theta_i-\theta_C(R_i)|}{\sqrt{\sigma_i^2+\sigma_{0,i}^2}}$ \\
\hline
\multicolumn{2}{c}{HMSFRs} \\
\multicolumn{2}{c}{($k=2.59$, $k_{0.05}=3.49$)} \\
G348.70-01.04 & $5.53$ \\
G023.44-00.18 & $4.01$ \\
G213.70-12.60 & $4.12$ \\
\multicolumn{2}{c}{HMSFRs+non-HMSFRs} \\
\multicolumn{2}{c}{Approach 1} \\
\multicolumn{2}{c}{($k=2.67$, $k_{0.05}=3.55$)} \\
G348.70-01.04 & $3.75$ \\
G110.10+02.47 & $4.87$ \\
G353.27+00.64 & $4.28$ \\
G173.72-02.70 & $4.35$ \\
G023.44-01.18 & $4.14$ \\
G045.37-00.22 & $4.24$ \\
G339.88-01.26 & $3.80$ \\
\multicolumn{2}{c}{Approach 2} \\
\multicolumn{2}{c}{($k=2.67$, $k_{0.05}=3.56$)} \\
G348.70-01.04 & $5.83$ \\
G110.10+02.47 & --- \\
G353.27+00.64 & $3.06$ \\
G213.70-12.60 & $4.00$ \\
G023.44-01.18 & $4.11$ \\
G045.37-00.22 & $4.13$ \\
\hline
\end{tabular}
\end{center}
\end{table}

\begin{table*}[]
\begin{center}
\caption{Solution with account for the natural azimuthal velocity dispersion after excluding from consideration the masers
with data outliers and the anomalous object}
\label{Tab4}
\medskip
\begin{tabular}{c|c|c|c}
\hline
Characteristic & HMSFRs & HMSFRs + non-HMSFRs & HMSFRs + non-HMSFRs \\
& & (approach 1) & (approach 2) \\
\hline
$N$ & 109 & 137 & 138\\
$\sigma_0$ or $\sigma_{0,1}$, km\,s$^{-1}$ & $3.85\pm 0.38$ & $6.56\pm 0.57$ & $4.34\pm0.38$ \\
$\sigma_{0,2}$, km\,s$^{-1}$ & & & $15.2\pm1.3$ \\
$q$ & $1^{+0}_{-0.018}$ & $0.943\pm 0.023$ & $1^{+0}_{-0.014}$ \\
$\kappa_1$, kpc$^{-1}$ & $0.1173\pm 0.0023$ & $0.1012\pm 0.0025$ & $0.0830\pm 0.0020$ \\
$\Phi_{0,1}$,~km$^2$\,s$^{-2}$ & $264.6\pm 6.0$ & $264.5\pm 7.2$ & $264.6\pm 5.3$ \\
$\alpha$ & $0.2495\pm 0.0046$ & $0.1651\pm 0.0046$ & $0.1068\pm 0.0031$ \\
$\kappa_2$,~kpc$^{-1}$ & $0.05455\pm 0.00085$ & $0.0575\pm 0.0011$ & $0.05645\pm 0.00055$ \\
$\Phi_{0,2}$,~km$^2$\,s$^{-2}$ & $321.2\pm 1.9$ & $318.9\pm 3.1$ & $315.6\pm 1.7$ \\
$\Phi_{0,3}$,~km$^2$\,s$^{-2}$ & $226.5\pm 9.7$ & $223.6\pm 13.4$ & $224.3\pm 7.4$ \\
$\kappa_3$,~kpc & $0.75\pm 0.12$ & $0.66\pm 0.25$ & $0.79\pm 0.12$ \\
\hline
\end{tabular}
\end{center}
\end{table*}

Parameter $z_0$ of the St\"ackel potential was determined from the following relation:
\begin{equation}
\displaystyle
z_0^2(R)
=\frac{\displaystyle 3\,\frac{\partial \Phi}{\partial R} (R,0)+
R\left[\displaystyle\frac{\partial^2 \Phi}{\partial R^2} (R,0)-4\,\frac{\partial^2 \Phi}{\partial z^2} (R,0)\right]}%
{\displaystyle\frac{\partial^3 \Phi}{\partial z^2 \partial R} (R,0)}
-R^2\,.
\end{equation}
Similar functions for $z_0$ were constructed by Einasto
and Rummel (1970), and also Osipkov (1975). We used the three-component potential from Gardner
et al. (2011) as potential $\Phi(R,z)$ to find $z_0$ since data on the vertical structure of the Galaxy were used in its
construction. The derived value $z_0=5.3$~kpc agrees with the estimates presented in other papers: $z_0=2.2 \div
7$~kpc (Hori, 1962; Kuzmin, 1953, 1956; Malasidze, 1973). We should note that some of these
stu\-dies use distances to the center of the Galaxy $R_0$, that are lower than the current estimate. For example, $R_0=7.0$ kpc in Kuzmin (1956) ($z_0=3.1$~kpc), $R_0=7.2$~kpc in Kuzmin (1953) ($z_0=3.6$~kpc). Additionally,
all the mentioned papers considered spherical models, thus decreasing~$z_0$. The value of $z_0$ derived by
us is close to $z_0=4.8$~kpc, reported by Malasidze (1973).

\section{\large{DISCUSSION}}

The options for the procedure of determining model parameters for the potential of the Galaxy from
masers considered in this work show that taking into account even a small natural azimuthal velocity dispersion
for masers is ne\-cessary, given the modern high-accuracy data on these objects. The derived azimuthal
dispersion for the joint sample $\sigma_0=6.56\pm 0.57$~km\,s$^{-1}$ is intermediary between the
radial ($\sigma_{U,0}=9.4\pm 0.9$~km\,s$^{-1}$) and vertical ($\sigma_{W,0}=5.9\pm 0.8$~km\,s$^{-1}$) natural dispersions found by
Rastorguev et al. (2017) using almost the same sample, i.e. it agrees with them. However the azimuthal dispersion
for the homogeneous HMSFR sample is significantly
lower: close values were derived by two methods~$\sigma_{0,1}=3.9\pm 0.4$ and $4.3\pm 0.4$~km\,s$^{-1}$
(Table~\ref{Tab4}). On the other hand, the dispersion for non-HMSFR masers $\sigma_{0,2}=15.2\pm1.3$~km\,s$^{-1}$ (even after excluding the anomalous object) is about~3.5--3.9 times higher than for HMSFRs, despite the expected general similarity
of these types, which not only confirms the kinematic inhomogeneity of non-HMSFRs compared to
HMSFRs, but implies a significant internal inhomogeneity of the non-HMSFR group. The small volume
of the non-HMSFR sample ($N=32$) and the large number of object types there (14 types) do not allow us
to determine the dispersion for individual classes.

That said, introducing two natural dispersions for the joint sample lead to a solution without significant
shifts of point parameter estimates compared to the results for the homogeneous HMSFR group, moreover,
with smaller statistical parameter errors on the whole. We therefore chose the model with two natural
dispersions as the final version for the St\"ackel generalization, derived after the exclusion of masers with
excessive residuals and the anomalous object (Table~\ref{Tab4}, approach~2; Fig.~\ref{Fig3}c).

The derived model rotation curves (Fig.~\ref{Fig3}) are
close to models constructed from masers at $R>3$--$4$~kpc in Rastorguev et al. (2017); Reid et al. (2019). Our model reconstructs more accurately the decline of the rotation curve in the outer part of the Galaxy than
the one in Reid et al. (2019), since the latter a priori
used a smoother function form for the model. The
model in Rastorguev et al. (2017) reveals more details
since they used a polynomial fourth-order model for
fitting. A good agreement is achieved in the common~$R$ interval and with the rotation curves based on classical
Cepheids in the recent works of Ablimit et al. (2020); Mroz et al. (2019), which also show a slow but
steady decreasing velocity trend at, $R>
6$~kpc. The same is also true for the rotation curve from red giants
(Eilers et al., 2019). On the whole, modern rotation
curves of the flat subsystems coincide rather well outside
of the central region of the Galaxy (see also Bland-Hawthorn and Gerhard (2016)). The halo circular
velocity curves derived from Gaia data on RR Lyrae stars (Wegg et al., 2019) are close to our model,
especially in the outer part of the galaxy. At $R< 3$~kpc our model cannot be reliable, if only due to the fact
that there it is based on only a few objects, to say nothing of the bar influence (Chemin et al., 2015), but it is
still sufficiently close to the rotation curve constructed
in the survey by Bland-Hawthorn and Gerhard (2016).

Dynamic modeling in this work was done in the assumption that the average rotation velocity of masers
is close to the circular velocity. This assumption agrees with the low azimuthal velocity dispersion which we
derived for masers, especially for HMSFRs. Direct estimates of the asymmetric shift for masers are represented
by $\overline{V_s}$ in Reid et al. (2019): they are located in the $-2\div-9$~km s $^{-1}$ interval for different versions of analysis, but even the most accurate of the estimates, $\overline{V_s}=-3.1\pm2.2$~km s $^{-1}$, does not differ significantly
from zero. Similar results were obtained also in Reid et al. (2014). Thus, the asymmetric shift can be neglected for masers, at least with the current level of accuracy.

\begin{figure*}
\includegraphics[scale=0.8]{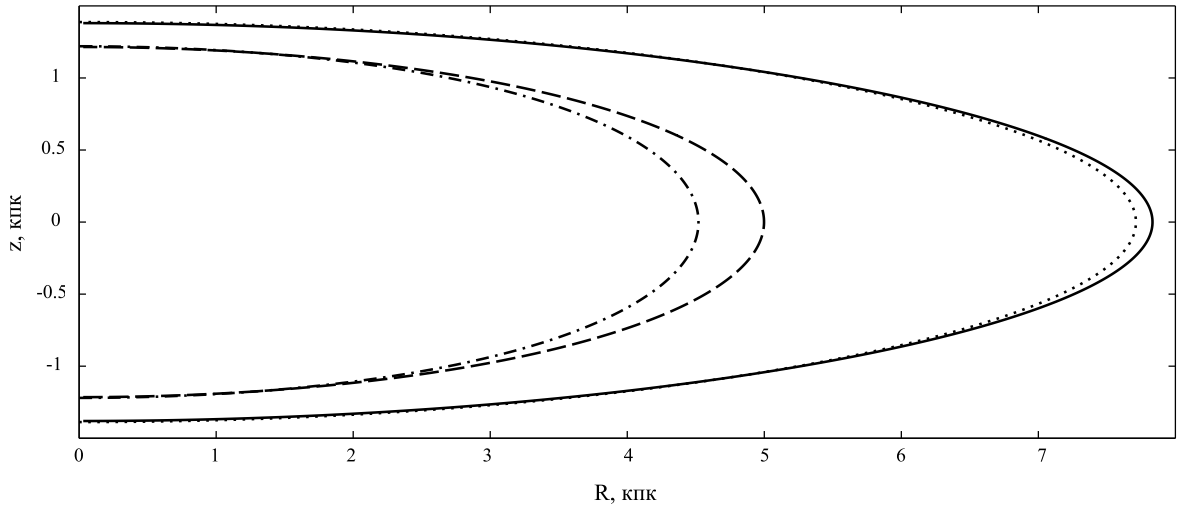}
\caption{Model equidensites for
$\rho=0.1$~$M_{\odot}$\,pc$^{-3}$. The solid
curve is the three-component model; the dashed curve
shows the halo; the dotted curve represents the disk, and
the dashed-and-dotted curve is the bulge.}
\label{Fig4}
\end{figure*}

We used Poisson’s equation to obtain an analytical expression for spatial density for the constructed
St\"ackel potential [(formulas \eqref{f4}, \eqref{f13}--\eqref{f16}] The corresponding equidensites (equal density curves) are presented
in Fig.~\ref{Fig4} for a density of $\rho=0.1$~$M_{\odot}$\,pc$^{-3}$. As is evident from the figure, the halo in the constructed
model is non-spherical. The model also does not adequately agree with the modern estimates of the vertical
scale of the thin ($300\pm50$ pc) and thick disk ($900\pm180$~pc)(Bland-Hawthorn and Gerhard, 2016).

The derived non-too-realistic equidensites raise the question of the necessity to additionally introduce
into the model a vertical distribution of density in one form or another when using Rodionov’s method. I.e.,
despite the interesting results of St\"ackel modeling in a series of theoretical studies (e.g., Kuzmin (1952);
Kuzmin and Malasidze (1987); \linebreak Osipkov (1975)), it turns out that the practical use of such methods in general
does not lead to a construction of models of acceptable adequacy. We should note that besides
some test work carried out by our group (Gromov et al. (2015, 2016)), there were no earlier attempts to use the
St\"ackel generalization in practice (applicable to real data). Such a result was therefore not obvious. Considering
the unrealistic equidensites obtained for disks in Binney and Wong (2017); Famaey and Dejonghe
(2003), where other methods of St\"ackel modeling were used (including the \flqq St\"ackel fudge\frqq), this problem is
of a more general nature and not merely a shortcoming of the St\"ackel generalization method.



Note that preliminary modeling based on the new catalog of Reid et al. (2019) gives similar results, i.e. the
nature of the vertical distribution is data insensitive.


Thus, the vertical structure of the model obtained in this work implies that using a direct St\"ackel generalization
method is insufficient for constructing realistic St\"ackel models. This would require solving the
question of accounting for the known properties of the vertical mass distribution in the Galaxy in St\"ackel
modeling. We considered several approaches and  their combinations as possibilities: redefining the function $\varphi(\xi)$ with account for the vertical distribution; introducing several additional conditions when estimating
parameters (for example, we were able to obtain an adequate disk thickness in the assumption of a barometric
vertical density distribution in preliminary calculations); varying the parameter $z_0$ in order to make
the shape of the halo more spherical. We propose to study these approaches in our next paper.

For the constructed model the solar neighborhood
density
($R_0=8.34$~kpc~\cite{Reid}) $\rho_{\odot}=0.082$~$M_{\odot}$\,pc$^{-3}$ is in good agreement with the current
estimates:
$\rho_{\odot}=0.08$--$0.11$~$M_{\odot}$\,pc$^{-3}$(Loktin and Marsakov, 2009), $\rho_{\odot}=0.097\pm0.013$~$M_{\odot}$\,pc$^{-3}$(Bland-Hawthorn and Gerhard, 2016).

The model mass in a sphere of 50 kpc radius is $M_{50}=0.82 \times 10^{12}~M_{\odot}$, which is somewhat higher than in the other studies. For example, Deason et al. (2012) found $M_{50}=(0.42\pm0.04) \times 10^{12}~M_{\odot}$, and Williams and Evans (2015)~--- $M_{50}=(0.45\pm0.15) \times 10^{12}~M_{\odot}$. The model mass in a 20~kpc sphere turned out equal to $0.35 \times 10^{12}~M_{\odot}$. The overestimated masses are probably also linked to omitting the data on the vertical
density distribution.

\section{\large{CONCLUSIONS}}

We constructed an analytical three component St\"ackel model of the Galaxy using the data on masers
with trigonometric parallaxes, proper motions, and radial velocities.

The maser data were used to estimate the parameters of the components --- halo, disk, and bulge --- for a
model representing the potential in the Galactic plane. We show that non-HMSFR masers, i.e., masers not
related to high-mass star forming regions (HMSFRs), are kinema\-tically inhomogeneous with respect to
HMSFR masers: the natural (non-instrumental) dispersion of their azimuthal velocity components,
$\sigma_{0,2}=15.2\pm1.3$~km\,s$^{-1}$ in 3.5--3.9 times higher than the similar dispersion $\sigma_{0,1}=(3.9\div 4.3)\pm 0.4$~km\,s$^{-1}$ for HMSFRs. This fact should be taken into account when using data on the kinematics of inhomogeneous
maser types. The final model was obtained with account for the difference in natural dispersions for
these two groups --- HMSFRs and non-HMSFRs. The model circular velocity curve is in good agreement with the observed data.

The derived model of the Galactic plane potential served as a basis for constructing a St\"ackel model of
the Galaxy using Rodionov’s method (Rodionov, 1974), which allows us to determine the values of the
potential in the entire space. The model gives close to real values of the density in the solar neighborhood of $\rho_{\odot}=0.082$~$M_{\odot}$\,pc$^{-3}$ and the total mass of $M_{50}=0.82 \times 10^{12}~M_{\odot}$ (accurate to an order of magnitude).
However, the nature of equidensites in the meridian plane of the model shows that direct use of
the St\"ackel generalization method alone is not enough to construct realistic St\"ackel models. To that end, one
needs to develop methods of directly accounting for data on the vertical density distribution in the Galaxy
when using Rodionov’s method.


Despite the uncovered but solvable complication, the approach on the whole has significant potential:
the proposed simple and, especially, analytical method of computing function
$\varphi(\xi)$ of the St\"ackel potential can be used when finding actions and may
significantly simplify the corresponding algorithms of constructing the phase model of the Galaxy.

The authors are grateful to V.V. Bobylev for providing
the information on three masers with full data and A.V. Veselova for the help with data reduction, and
especially for the complex task of identifying object types. The authors are grateful to the anonymous referee
for useful comments.

%


\section{\large{FUNDING}}
The reported study was funded by RFBR, project number 19-32-90144.

\section{\large{CONFLICT OF INTEREST}}
The authors declare no conflict of interest.


\begin{thebibliography}{19}

\bibitem{Kuzm1}
I. Ablimit, G. Zhao, C. Flynn, and S. A. Bird, Astrophys.
J. {\bf 895} (1), L12 (2020).


\bibitem{Malas}
J. Binney, Monthly Notices Royal Astron. Soc.  {\bf 401} (4),
2318 (2010).


\bibitem{Osip1}
J. Binney, Monthly Notices Royal Astron. Soc. {\bf 426} (2),
1324 (2012).

\bibitem{BinMer}
J. Binney and M. Merrifield,  {\it Galactic Astronomy}
(Princeton University Press, Princeton, NJ, 1998).

\bibitem{Bond+10}
J. Binney and L. K. Wong, Monthly Notices Royal
Astron. Soc. {\bf 467} (2), 2446 (2017).

\bibitem{Rastor}
J. Binney, R. L. Davies, and G. D. Illingworth,
Astrophys. J. {\bf 361}, 78 (1990).

\bibitem{Ogor}
J. Bland-Hawthorn and O. Gerhard, Annual Rev.
Astron. Astrophys. {\bf 54}, 529 (2016).

\bibitem{Henon}
N. A. Bond, Z. Ivezic, B. Sesar, et al., Astrophys. J. {\bf 716}
(1), 1 (2010).


\bibitem{Rod1}
R. A. Burns, T. Handa, H. Imai, et al., Monthly Notices
Royal Astron. Soc. {\bf 467} (2), 2367 (2017).

\bibitem{Binney4}
L. Chemin, F. Renaud, and C. Soubiran, Astron. and
Astrophys. {\bf 578}, A14 (2015).


\bibitem{Merif}
J. O. Chibueze, H. Hamabata, T. Nagayama, et al.,
Monthly Notices Royal Astron. Soc. {\bf 466} (4), 4530
(2017).


\bibitem{Stack}
J. O. Chibueze, H. Sakanoue, T. Nagayama, et al.,
Publ. Astron. Soc. Japan {\bf 66} (6), 104 (2014).

\bibitem{Edin}
A. J. Deason, V. Belokurov, N. W. Evans, and J. An,
Monthly Notices Royal Astron. Soc. {\bf 424} (1), L44
(2012).

\bibitem{Deh}
W. Dehnen, Astron. J. {\bf 118} (3), 1201 (1999).

\bibitem{Posti}
A. S. Eddington, Monthly Notices Royal Astron. Soc.
{\bf 76}, 37 (1915).

\bibitem{Binney2}
A.-C. Eilers, D. W. Hogg, H.-W. Rix, and M. K. Ness,
Astrophys. J. {\bf 871} (1), 120 (2019).

\bibitem{Binney3}
J. Einasto and U. Rummel, Astrophysics {\bf 6} (2), 120
(1970).

\bibitem{Sand}
B. Famaey and H. Dejonghe, Monthly Notices Royal
Astron. Soc. {\bf 340} (3), 752 (2003).

\bibitem{Binney1}
E. Gardner, P. Nurmi, C. Flynn, and S. Mikkola,
Monthly Notices Royal Astron. Soc. {\bf 411} (2), 947
(2011).

\bibitem{Sanders}
A. O. Gromov, I. I. Nikiforov, and L. P. Ossipkov, Baltic
Astronomy {\bf 24}, 150 (2015).

\bibitem{Satoh}
A. O. Gromov, I. I. Nikiforov, and L. P. Ossipkov, Baltic
Astronomy {\bf 25}, 53 (2016).

\bibitem{Lok}
M. Henon and C. Heiles, Astron. J. {\bf 69}, 73 (1964).

\bibitem{Hawt}
L. Hernquist, Astrophys. J. {\bf 356}, 359 (1990).

\bibitem{Fam}
G. Hori, Publ. Astron. Soc. Japan {\bf 14}, 353 (1962).

\bibitem{Binney5}
M. Kaasalainen and J. Binney, Monthly Notices Royal
Astron. Soc. {\bf 268}, 1033 (1994).

\bibitem{Gr1}
G. G. Kuzmin, Publ. Tartu Astrofiz. Obs. {\bf 32}, 332
(1952).

\bibitem{Gr2}
G. G. Kuzmin, Izvestiya Akad. Nauk ESSR {\bf 2}, 3 (1953).

\bibitem{Kuzm2}
G. G. Kuzmin, Astron. Zh. {\bf 33}, 27 (1956).

\bibitem{Kuzm6}
G. G. Kuzmin and G. A. Malasidze, Akademiia Nauk
Gruzinskoj SSR {\bf 54}, 565 (1969).

\bibitem{Hen}
G. G. Kuzmin and G. A. Malasidze, Publ. Tartu Astrofiz.
Obs. {\bf 52}, 48 (1987).

\bibitem{NV18b}
G. G. Kuzmin, U. I. K. Veltmann, and P. L. Tenjes,
Publ. Tartu Astrofiz. Obs. {\bf 51}, 232 (1986).

\bibitem{Reid}
A. V. Loktin and Marsakov V. A., {\it Lecscii po zvezdnoi astronomii}
(SFedU, Rostov-na-Dony, 2009).

\bibitem{Chibueze+17}
G. A. Malasidze, Proc. conf. on {\it Dynamics of Galaxies
and Star Clusters}, pp. 93–98 (1973).

\bibitem{Burns+17}
M. R. Merrifield, Astron. J. {\bf 102}, 1335 (1991).

\bibitem{N12}
P. Mroz, A. Udalski, D. M. Skowron, et al., Astrophys. J.
{\bf 870} (1), L10 (2019).

\bibitem{Chibueze}
I. I. Nikiforov, Astron. and Astrophys. Transact. {\bf 27} (3),
537 (2012).

\bibitem{Einasto}
I. I. Nikiforov and A. V. Veselova, Astr. Lett. {\bf 44} (11),
699 (2018).

\bibitem{Gard}
K. F. Ogorodnikov, {\it Dinamika zvezdnykh sistem} (Gosydarstvennoe
izdatel’stvo fiziko-matematichskoi literatury,
Moskva, 1958) [in Russian].

\bibitem{Kuzm3}
L. P. Osipkov, in {\it Vestnik Leningradskogo Universiteta.
Seriya 1. Matematika. Mekhanika. Astronomiia},
pp. 151–158 (1975) [in Russian].

\bibitem{Kuzm4}
L. Posti, J. Binney, C. Nipoti, and L. Ciotti, Monthly
Notices Royal Astron. Soc. {\bf 447} (4), 3060 (2015).

\bibitem{Hori}
A. S. Rastorguev, N. D. Utkin, M. V. Zabolotskikh,
et al., Astrophysical Bulletin {\bf 72} (2), 122 (2017).

\bibitem{Malas1}
M. J. Reid, K. M. Menten, A. Brunthaler, et al., Astrophys.
J. {\bf 885} (2), 131 (2019).

\bibitem{Reid2}
V. I. Rodionov, in {\it Vestnik Leningradskogo Universiteta.
Seriya 1. Matematika. Mekhanika. Astronomiia},
pp. 142–148 (1974) [in Russian].

\bibitem{Mroz+19}
J. Sanders, Monthly Notices Royal Astron. Soc.
{\bf 426} (1), 128 (2012).

\bibitem{Ablimit+20}
J. Sanders and J. Binney, Monthly Notices Royal
Astron. Soc. {\bf 457} (2), 2107 (2016).

\bibitem{Eilers+19}
M. J. Reid, K. M. Menten, A. Brunthaler, et al., Astrophys.
J. {\bf 793} (2), 72 (2014).

\bibitem{Wegg+19}
C. Satoh and M. Miyamoto, Publ. Astron. Soc. Japan
{\bf 28} (4), 599 (1976).

\bibitem{Chemin+15}
P. St\"ackel, Math. Annal. {\bf 35}, 91 (1890).

\bibitem{Deas}
C. Wegg, O. Gerhard, and M. Bieth, Monthly Notices
Royal Astron. Soc. {\bf 485} (3), 3296 (2019).

\bibitem{Wil}
A. A. Williams and N. W. Evans, Monthly Notices Royal
Astron. Soc. {\bf 454} (1), 698 (2015).

\end{thebibliography}
\end{document}